\documentclass[12pt,twoside]{article}
\usepackage[T1]{fontenc}
\usepackage[latin9]{inputenc}
\usepackage[letterpaper]{geometry}
\usepackage{amsmath}
\begin{document}
\setcounter{page}{0} \begin{titlepage} \vspace{0.5cm}

\begin{center}
{\large On the multiparametric ${\cal {U}}_{q}[D_{n+1}^{(2)}]$ vertex
model}\\
 \vspace{1.5cm}
 {\large Ricardo S. Vieira}%
\footnote{rsvieira@df.ufscar.br%
}{\large{} and A. Lima-Santos}%
\footnote{dals@df.ufscar.br%
} \\
 \vspace{0.5cm}
 {\em Universidade Federal de São Carlos, Departamento de Física
\\
 C.P. 676, 13565-905~~São Carlos(SP), Brasil}\\
 
\par\end{center}

\vspace{1.5cm}

\begin{abstract}
In this paper we consider families of multiparametric $R$-matrices
to make a systematic study of the boundary Yang-Baxter equations in
order to discuss the corresponding families of multiparametric $K$-matrices.
Our results are indeed non-trivial generalization of the $K$-matrix
solutions of the ${\cal {U}}_{q}[D_{n+1}^{(2)}]$ vertex model when
distinct reflections and extra free-parameters are admissible.
\end{abstract}
\vspace{2cm}
 \centerline{PACS numbers: 05.50+q, 02.30.} \vspace{0.1cm}
 \centerline{Keywords: Integrable lattice models, boundary Yang-Baxter
equations} \vfill{}
\centerline{\today} \vspace{1.5cm}
\end{titlepage}

\section{Introduction\label{sec:Intro}}

The Yang-Baxter equation is an operator relation for a matrix $R_{ij}(x)$
defined on the tensor product of two $N$-dimensional vectors spaces
$V_{i}$ and $V_{j}$, which reads 
\begin{equation}
R_{12}(x_{1})R_{13}(x_{1}x_{2})R_{23}(x_{2})=R_{23}(x_{2})R_{13}(x_{1}x_{2})R_{12}(x_{1}),\label{yb}
\end{equation}
where $x_{a}$ are arbitrary multiplicative spectral parameters.

Search for solutions of the Yang-Baxter equation is a important subject
in the studies of exactly solvable models. Actually, a $R$-matrix
satisfying the Yang-Baxter equation generates the Boltzmann weights
of an vertex model \cite{BAX} or the factorisable scattering amplitudes
between particles in relativistic field theories \cite{ZAM}.

A classification of the solutions of the Yang-Baxter equation for
$R$-matrices (with an extra $q$-parameter) was performed by $q$-deformation
in a given non-exceptional Lie algebra ${\cal {G}}$ \cite{FA,DRI,JIQ}.
This quantum group approach permits us to reduce the problem (\ref{yb})
to a linear one, in order to associate a fundamental trigonometric
$R$-matrix to each Lie algebra \cite{BAZ,JI} or Lie superalgebra
\cite{BAZ1,Zhang,WM}. \ 

The physical understanding of vertex models includes necessarily the
exact diagonalization of their transfer matrices, which can provide
us information about the free energy behavior and on the nature of
the elementary excitations. This step has been successfully achieved
for standard Lie algebras either by the analytical Bethe ansatz \cite{RE},
a phenomenological technique yielding us solely the transfer matrix
eigenvalues, or through the quantum inverse scattering method \cite{QISM,DEV,Kulish,MA},
an algebraic technique which gives us also the eigenvectors. Usually
these systems are studied with periodic boundary conditions but more
general boundaries can also be included in this framework as well.
\ Physical properties associated with the bulk of the system are
not expected to be influenced by boundary conditions in the thermodynamical
limit. Nevertheless, there are surface properties such as the interfacial
tension where the boundary conditions are of relevance. Moreover,
the conformal spectra of lattice models at criticality can be modified
by the effect of boundaries \cite{CAR1}.

Integrable systems with open boundary conditions can also be accommodated
within the framework of the Quantum Inverse Scattering Method \cite{SK}.
In addition to the solution of the Yang-Baxter equation governing
the dynamics of the bulk there is another fundamental ingredient,
the reflection matrices \cite{CHER}. These matrices, also referred
as $K$-matrices, represent the interactions at the boundaries and
compatibility with the bulk integrability requires these matrices
to satisfy the so-called reflection equations \cite{SK,CHER}.

The original formalism of \cite{SK} was extended to more general
systems in \cite{MEZ}, where is assumed that a regular $R$-matrix
satisfying the properties PT-symmetry, unitarity and crossing unitarity,
one can derive an integrable open chain Hamiltonian with the right
boundaries determined by the solutions of the reflection equation,
\begin{equation}
R_{21}(x/y)K_{2}^{-}(x)R_{12}(xy)K_{1}^{-}(y)=K_{1}^{-}(y)R_{21}(xy)K_{2}^{-}(x)R_{12}(x/y),\label{re}
\end{equation}
A similar equation should also hold for the $K^{+}$-matrix at the
opposite boundary. When these properties are fulfilled one can follow
the scheme devised in \cite{MEZ} and the matrix $K^{-}(x)$ is obtained
by solving the Eq.(\ref{re}) while the matrix $K^{+}(x)$ can be
obtained from the isomorphism 
\begin{equation}
K^{-}(x)\mapsto K^{+}(x)^{{\rm t}}=K^{-}(x^{-1}\eta^{-1})V^{{\rm t}}V.\label{auto}
\end{equation}
Here ${\rm t}$ stands for the transposition, $\eta$ is a crossing
parameter and $V$ a crossing matrix, both being specific to each
model.

At the moment, the study of general regular solutions of the reflection
equations for vertex models based on $q$-deformed Lie algebras \cite{BAZ,JI}
has been successfully accomplished. See \cite{LIM1} for instance
and references therein.

In reference \cite{WM1}, families of $R$-matrices not previously
foreseen by the framework of quantum groups was obtained by the baxterization
of the representations of the Birman-Wenzl-Murakami algebra. In this
paper we are taking into account these results in order to consider
the systems which were presented as rather non-trivial extensions
of Jimbo's ${\cal U}_{q}[D_{n+1}^{(2)}]$ $R$-matrix. The main interest
of this work is to present families of multiparametric $K$-matrix
associated with the multiparametric ${\cal U}_{q}[D_{n+1}^{(2)}]$
vertex model, by solving its functional reflection equations. $K$-matrices
for the Jimbo's ${\cal U}_{q}[D_{n+1}^{(2)}]$ vertex model was already
derived in \cite{LIM2}. Here we re-derive these results and also
present new solutions for the Jimbo's ${\cal U}_{q}[D_{n+1}^{(2)}]$
vertex model as well. 

This paper has been organized as follows. In Section \ref{sec:R}
we present the $R$-matrices which turn out to be highly non-trivial
extensions of the ${\cal U}_{q}[D_{n+1}^{(2)}]$ vertex model, as
presented in \cite{WM1}. In Section \ref{sec:K} we present with
some details the corresponding matrix elements of the $K$-matrix
solutions. Our conclusions are summarized in Section \ref{sec:KC}.
In Appendix $A$ we present the solutions with a block diagonal structure
compatible with the $n+1$ distinct $U(1)$ conserved charges of the
${\cal U}_{q}[D_{n+1}^{(2)}]$ vertex model \cite{GM}. In Appendix
$B$ we present a more general solution for the ${\cal U}_{q}[D_{2}^{(2)}]$
vertex model. In Appendix $C$ we present some special solutions.

\section{The multiparametric ${\cal U}_{q}[D_{n+1}^{(2)}]$ vertex model\label{sec:R}}

In what follows we will to consider the multiparametric ${\cal U}_{q}[D_{n+1}^{(2)}]$
$R$-matrix as derived in \cite{WM1} from the representations of
a quotient of the braid-monoid algebra denominated Birman-Wenzl-Murakami
algebra \cite{BW,Mura}. In this situation, the link between braid
algebra and the Yang-Baxter equation is made with the help of the
baxterization procedure and the permutator, 
\begin{equation}
P=\sum_{i,j=1}^{N}{\rm e}_{i,j}\otimes{\rm e}_{j,i},\label{perm}
\end{equation}
where $N=2n+2$ and ${\rm e}_{i,j}$ denotes the standard $N\times N$
Weyl matrices.

In fact, by defining a new matrix ${\check{R}}_{ab}(x)=P_{ab}R_{ab}(x)$
one can rewrite the Yang-Baxter equation in a form 
\begin{equation}
{\check{R}}_{12}(x_{1}){\check{R}}_{23}(x_{1}x_{2}){\check{R}}_{12}(x_{2})={\check{R}}_{23}(x_{2}){\check{R}}_{12}(x_{1}x_{2}){\check{R}}_{23}(x_{1}),\label{ybr}
\end{equation}
and its solution can be rewritten as follows, 
\begin{eqnarray}
\check{R}(x) & = & {\displaystyle \sum_{i\neq n+1,n+2}}(x^{2}-\zeta^{2})\left(x^{2}-q^{2}\right){\rm e}_{i,i}\otimes{\rm e}_{i,i}+q(x^{2}-1)(x^{2}-\zeta^{2}){\displaystyle \sum_{\substack{i\neq j,j^{\prime}\\
i,j\neq n+1,n+2
}
}}{\rm e}_{j,i}\otimes{\rm e}_{i,j}\nonumber \\
 & + & \frac{1}{2}q(x^{2}-1)(x^{2}-\zeta^{2})\sum_{\substack{i\neq j,j^{\prime}\\
j=n+1,n+2
}
}[(1+\kappa)\left({\rm e}_{j,i}\otimes{\rm e}_{i,j}+{\rm e}_{i,j}\otimes{\rm e}_{j,i}\right)\nonumber \\
 & + & (1-\kappa)\left({\rm e}_{j,i}\otimes{\rm e}_{i,j^{\prime}}+{\rm e}_{i,j}\otimes{\rm e}_{j^{\prime},i}\right)]+\sum_{i,j\neq n+1,n+2}g_{i,j}(x){\rm e}_{i^{\prime},j}\otimes{\rm e}_{i,j^{\prime}}\nonumber \\
 & - & (q^{2}-1)(x^{2}-\zeta^{2})\left[\sum_{\substack{i<j,i\neq j^{\prime}\\
i,j\neq n+1,n+2
}
}+x^{2}\sum_{\substack{i>j,i\neq j^{\prime}\\
i,j\neq n+1,n+2
}
}\right]{\rm e}_{j,j}\otimes{\rm e}_{i,i}\nonumber \\
 & - & \frac{1}{2}(q^{2}-1)(x^{2}-\zeta^{2})[(x+1)\left(\sum_{\substack{i<n+1\\
j=n+1,n+2
}
}+x\sum_{\substack{i>n+2\\
j=n+1,n+2
}
}\right)\left({\rm e}_{j,j}\otimes{\rm e}_{i,i}+{\rm e}_{i^{\prime},i^{\prime}}\otimes{\rm e}_{j^{\prime},j}\right)\nonumber \\
 & + & (x-1)\left(-\sum_{\substack{i<n+1\\
j=n+1,n+2
}
}+x\sum_{\substack{i>n+2\\
j=n+1,n+2
}
}\right)\left({\rm e}_{j^{\prime},j}\otimes{\rm e}_{i,i}+{\rm e}_{i^{\prime},i^{\prime}}\otimes{\rm e}_{j^{\prime},j^{\prime}}\right)]\nonumber \\
 & + & \frac{1}{2}\sum_{\substack{i\neq n+1,n+2\\
j=n+1,n+2
}
}\left[b_{i}^{+}(x)\left({\rm e}_{i^{\prime},j}\otimes{\rm e}_{i,j^{\prime}}+{\rm e}_{j,i^{\prime}}\otimes{\rm e}_{j^{\prime},i}\right)+b_{i}^{-}(x)\left({\rm e}_{i^{\prime},j}\otimes{\rm e}_{i\; j}+{\rm e}_{j,i^{\prime}}\otimes{\rm e}_{j,i}\right)\right]\nonumber \\
 & + & \sum_{i=n+1,n+2}\left[c_{i}^{+}(x){\rm e}_{i^{\prime},i}\otimes{\rm e}_{i,i^{\prime}}+c_{i}^{-}(x){\rm e}_{i,i}\otimes{\rm e}_{i,i}+d_{i}^{+}(x){\rm e}_{i^{\prime},i^{\prime}}\otimes{\rm e}_{i,i}+d_{i}^{-}(x){\rm e}_{i,i^{\prime}}\otimes{\rm e}_{i,i^{\prime}}\right],\nonumber \\
\label{Rhat}
\end{eqnarray}
where $i^{\prime}=N+1-i$ and $\zeta=q^{n}$. The respective Boltzmann
weights $g_{i,j}(x)$, $b_{i}^{\pm}(x)$, $c_{\nu}^{\pm}(x)$ and
$d_{\nu}^{\pm}(x)$ are given by
\begin{equation}
g_{i,j}(x)=\left\{ \begin{array}{c}
(x^{2}-1)\left[(x^{2}-\zeta^{2})+x^{2}(q^{2}-1)\right]\hfill i=j,\\
(q^{2}-1)\left[\zeta^{2}(x^{2}-1)q^{t_{i}-t_{j}}-\delta_{i,j^{\prime}}(x^{2}-\zeta^{2})\right]\hfill i<j,\\
(q^{2}-1)x^{2}\left[(x^{2}-1)q^{t_{i}-t_{j}}-\delta_{i,j^{\prime}}(x^{2}-\zeta^{2})\right]\quad\quad\hfill i>j,
\end{array}\right.
\end{equation}
\begin{equation}
b_{i}^{\pm}(x)=\left\{ \begin{array}{c}
\pm q^{\widetilde{t}_{i}}(q^{2}-1)(x^{2}-1)(x\pm\zeta)\hfill i<n+1,\\
q^{\widetilde{t}_{i}}(q^{2}-1)(x^{2}-1)x(x\pm\zeta)\quad\quad\hfill i>n+2,
\end{array}\right.
\end{equation}
\begin{equation}
c_{\nu}^{\pm}(x)=\pm\frac{1}{2}(q^{2}-1)(\zeta+1)x(x\mp1)(x\pm\zeta)+\frac{1}{2}(1+\nu\kappa)q(x^{2}-1)(x^{2}-\zeta^{2}),
\end{equation}
\begin{equation}
d_{\nu}^{\pm}(x)=\pm\frac{1}{2}(q^{2}-1)(\zeta-1)x(x\pm1)(x\pm\zeta)+\frac{1}{2}(1-\nu\kappa)q(x^{2}-1)(x^{2}-\zeta^{2}),
\end{equation}
where $\kappa=\pm1$ and the lower index $\nu=\pm1$ in the weights
$c_{\nu}^{\pm}(x)$ and $d_{\nu}^{\pm}(x)$ indicates the two possible
families of models. The explicit expressions for the variables $t_{i}$
and $\widetilde{t}_{i}$ are
\begin{equation}
t_{i}=\left\{ \begin{array}{c}
i+1\qquad\hfill i<n+1,\\
n+\frac{3}{2}\qquad\hfill i=n+1,n+2,\\
i-1\qquad\hfill i>n+2,
\end{array}\right.\label{ntildet}
\end{equation}
\begin{equation}
\widetilde{t}_{i}=\left\{ \begin{array}{c}
i+\frac{1}{2}\qquad\hfill i<n+1,\\
i-n-\frac{5}{2}\qquad\hfill i>n+2.
\end{array}\right.\label{tildet}
\end{equation}

As noted in \cite{WM1} it is not difficult to recognize that expressions
(\ref{Rhat}-\ref{tildet}) for the branch $\kappa=1$ and $\nu=1$
indeed reproduce the usual ${\cal U}_{q}[D_{n+1}^{(2)}]$ $R$-matrix
\cite{JI}. This means that in general the $R$-matrix should be considered
as a non-trivial generalization of Jimbo's ${\cal U}_{q}[D_{n+1}^{(2)}]$
vertex model.

Now, we would like to present some useful properties satisfied by
the $R$-matrix $R(x)=P\check{R}(x)$ where $\check{R}(x)$ refers
to the matrix given in Eq.(\ref{Rhat}). Besides regularity and unitarity
this $R$-matrix satisfies the so-called $PT$ symmetry given by 
\begin{equation}
R_{21}(x)=P_{12}R_{12}(x)P_{12}=[R_{12}]^{t_{1}t_{2}}(x),\label{PT}
\end{equation}
where the symbol $t_{k}$ denotes the transposition in the space with
index $k$. Yet another property is the crossing symmetry, namely
\begin{equation}
R_{12}(x)=\frac{\rho(x)}{\rho(\zeta/x)}V_{1}[R_{12}]^{t_{2}}(\zeta/x)V_{1}^{-1},\label{CRO}
\end{equation}
where $\rho(x)$ is a convenient normalization 
\begin{equation}
\rho(x)=q(x^{2}-1)(x^{2}-\zeta^{2}),
\end{equation}
while the only non-null entries of the normalized matrix $V$ are
the minor diagonal elements $V_{i,i^{\prime}}$, namely
\begin{equation}
V_{i,i^{\prime}}=\left\{ \begin{array}{c}
q^{i-1}\qquad\hfill i<n+1,\\
q^{n-\frac{1}{2}}\qquad\hfill i=n+1,n+2,\\
q^{i-3}\qquad\hfill i>n+2.
\end{array}\right.\label{vmat}
\end{equation}

\section{The multiparametric ${\cal U}_{q}[D_{n+1}^{(2)}]$ $K$-matrix\label{sec:K}}

The purpose of this work is to investigate the general families of
regular solutions of the reflection equation Eq.(\ref{re}). Regular
solutions mean that the $K$-matrices have the general form 
\begin{equation}
K^{-}(x)=\sum_{i,j=1}^{N}k_{i,j}(x)\; e_{i,j},\label{KM}
\end{equation}
such that the condition $k_{i,j}(1)=\delta_{i,j}$ holds for all matrix
elements.

The direct substitution of the $K$ and the $R$ matrices in the reflection
equation, leave us with a system of $N^{4}$ functional equations
for the entries $k_{i,j}(x)$. In order to solve these equations we
shall make use of the derivative method. Thus, by differentiating
the equation Eq.(\ref{re}) with respect to $y$ and setting $y=1$,
we obtain a set of algebraic equations for the matrix elements $k_{i,j}$
involving the single variable $x$ and $N^{2}$ parameters 
\begin{equation}
\beta_{i,j}=\left.\frac{dk_{i,j}(y)}{dy}\right|_{y=1}\qquad i,j=1,2,...,N.\label{param.}
\end{equation}
Although we obtain a large number of equations only a few of them
are actually independent and a direct inspection of those equations,
in the lines described in \cite{LIM2}, allows us to find the branches
of regular solutions. In what follows we shall present our findings
for the regular solutions of the reflection equation associated with
the multiparametric ${\cal U}_{q}[D_{n+1}^{(2)}]$ vertex model. The
special cases are presented in the appendices.

\subsection{Non-diagonal matrix elements\label{sub:Knd}}

All families of solutions have a common structure for its non-diagonal
matrix elements $k_{i,j}(x)$ with $i\neq j$ and different of $k_{n+1,n+2}(x)$
and $k_{n+2,n+1}(x)$. The minor diagonal elements are given by
\begin{eqnarray}
k_{i,i^{\prime}}(x) & = & q^{t_{i}-2n}\Gamma(n)^{2}\left(\frac{\beta_{1,i^{\prime}}}{\beta_{1,N}}\right)^{2}k_{1,N}(x),\nonumber \\
\qquad i & \neq & 1,n+1,n+2,N,\label{nd.1}
\end{eqnarray}
\begin{equation}
k_{N,1}(x)=q^{2n-3}\left(\frac{\beta_{2,1}}{\beta_{1,N-1}}\right)^{2}k_{1,N}(x).\label{nd.2}
\end{equation}
The elements of the first row are given by
\begin{eqnarray}
k_{1,j}(x) & = & \left(\frac{\beta_{1,j}}{\beta_{1,N}}\right)G(x)k_{1,N}(x),\nonumber \\
\qquad j & \neq & n+1,n+2,\label{nd.3}
\end{eqnarray}
\begin{equation}
k_{1,n+1}(x)=\left(\frac{\beta_{+}+x\beta_{-}}{\beta_{1,N}}\right)G(x)k_{1,N}(x),\label{nd.4}
\end{equation}
\begin{equation}
k_{1,n+2}(x)=\left(\frac{\beta_{+}-x\beta_{-}}{\beta_{1,N}}\right)G(x)k_{1,N}(x).\label{nd.5}
\end{equation}
The elements of the first column are given by
\begin{eqnarray}
k_{i,1}(x) & = & q^{t_{i}-3}\left(\frac{\beta_{2,1}}{\beta_{1,N-1}}\right)\left(\frac{\beta_{1,i^{\prime}}}{\beta_{1,N}}\right)G(x)k_{1,N}(x),\nonumber \\
\qquad i & \neq & n+1,n+2,\label{nd.6}
\end{eqnarray}
\begin{equation}
k_{n+1,1}(x)=q^{t_{n+1}-3}\left(\frac{\beta_{2,1}}{\beta_{1,N-1}}\right)\left(\frac{\beta_{+}-\kappa\epsilon x\beta_{-}}{\beta_{1,N}}\right)G(x)k_{1,N}(x),\label{nd.7}
\end{equation}
\begin{equation}
k_{n+2,1}(x)=q^{t_{n+2}-3}\left(\frac{\beta_{2,1}}{\beta_{1,N-1}}\right)\left(\frac{\beta_{+}+\kappa\epsilon x\beta_{-}}{\beta_{1,N}}\right)G(x)k_{1,N}(x).\label{nd.8}
\end{equation}
The elements of the last row are given by
\begin{eqnarray}
k_{N,j}(x) & = & \epsilon q^{n-2}\left(\frac{\beta_{2,1}}{\beta_{1,N-1}}\right)\left(\frac{\beta_{1,j}}{\beta_{1,N}}\right)x^{2}G(x)k_{1,N}(x),\nonumber \\
\qquad j & \neq & n+1,n+2,\label{nd.9}
\end{eqnarray}
\begin{equation}
k_{N,n+1}(x)=\epsilon q^{n-2}\left(\frac{\beta_{2,1}}{\beta_{1,N-1}}\right)\left(\frac{x\beta_{+}+\epsilon q^{n}\beta_{-}}{\beta_{1,N}}\right)xG(x)k_{1,N}(x),\label{nd.11}
\end{equation}
\begin{equation}
k_{N,n+2}(x)=\epsilon q^{n-2}\left(\frac{\beta_{2,1}}{\beta_{1,N-1}}\right)\left(\frac{x\beta_{+}-\epsilon q^{n}\beta_{-}}{\beta_{1,N}}\right)xG(x)k_{1,N}(x).\label{nd.12}
\end{equation}
The elements of the last column
\begin{eqnarray}
k_{i,N}(x) & = & \epsilon q^{t_{i}-n-2}\left(\frac{\beta_{1,i^{\prime}}}{\beta_{1,N}}\right)x^{2}G(x)k_{1,N}(x),\nonumber \\
\qquad i & \neq & n+1,n+2,\label{nd.13}
\end{eqnarray}
\begin{equation}
k_{n+1,N}(x)=\epsilon q^{t_{n+1}-n-2}\left(\frac{x\beta_{+}-\kappa q^{n}\beta_{-}}{\beta_{1,N}}\right)xG(x)k_{1,N}(x),\label{nd.14}
\end{equation}
\begin{equation}
k_{n+2,N}(x)=\epsilon q^{t_{n+2}-n-2}\left(\frac{x\beta_{+}+\kappa q^{n}\beta_{-}}{\beta_{1,N}}\right)xG(x)k_{1,N}(x).\label{nd.15}
\end{equation}
Moreover, the remained non-diagonal matrix elements above the minor
diagonal are given by
\begin{eqnarray}
k_{i,j}(x) & = & q^{t_{i}-n-1}\Gamma(n)\left(\frac{\beta_{1,i^{\prime}}}{\beta_{1,N}}\right)\left(\frac{\beta_{1,j}}{\beta_{1,N}}\right)G(x)k_{1,N}(x),\nonumber \\
i^{\prime} & > & j,\label{nd.16}
\end{eqnarray}
\begin{equation}
k_{n+1,j}(x)=q^{t_{n+1}-n-1}\Gamma(n)\left(\frac{\beta_{1,j}}{\beta_{1,N}}\right)\left(\frac{\beta_{+}-\kappa\epsilon x\beta_{-}}{\beta_{1,N}}\right)G(x)k_{1,N}(x),\label{nd.17}
\end{equation}
\begin{equation}
k_{n+2,j}(x)=q^{t_{n+2}-n-1}\Gamma(n)\left(\frac{\beta_{1,j}}{\beta_{1,N}}\right)\left(\frac{\beta_{+}+\kappa\epsilon x\beta_{-}}{\beta_{1,N}}\right)G(x)k_{1,N}(x),\label{nd.18}
\end{equation}
\begin{equation}
k_{i,n+1}(x)=q^{t_{i}-n-1}\Gamma(n)\left(\frac{\beta_{1,i^{\prime}}}{\beta_{1,N}}\right)\left(\frac{\beta_{+}+x\beta_{-}}{\beta_{1,N}}\right)G(x)k_{1,N}(x),\label{nd.19}
\end{equation}
\begin{equation}
k_{i,n+2}(x)=q^{t_{i}-n-1}\Gamma(n)\left(\frac{\beta_{1,i^{\prime}}}{\beta_{1,N}}\right)\left(\frac{\beta_{+}-x\beta_{-}}{\beta_{1,N}}\right)G(x)k_{1,N}(x),\label{nd.20}
\end{equation}
while the remained non-diagonal matrix elements below the minor diagonal
are given by
\begin{eqnarray}
k_{i,j}(x) & = & \epsilon q^{t_{i}-2n-1}\Gamma(n)\left(\frac{\beta_{1,i^{\prime}}}{\beta_{1,N}}\right)\left(\frac{\beta_{1,j}}{\beta_{1,N}}\right)x^{2}G(x)k_{1,N}(x),\nonumber \\
i^{\prime} & < & j,\label{nd.21}
\end{eqnarray}
\begin{equation}
k_{n+1,j}(x)=\epsilon q^{t_{n+1}-2n-1}\Gamma(n)\left(\frac{\beta_{1,j}}{\beta_{1,N}}\right)\left(\frac{x\beta_{+}-\kappa q^{n}\beta_{-}}{\beta_{1,N}}\right)xG(x)k_{1,N}(x),\label{nd.22}
\end{equation}
\begin{equation}
k_{n+2,j}(x)=\epsilon q^{t_{n+2}-2n-1}\Gamma(n)\left(\frac{\beta_{1,j}}{\beta_{1,N}}\right)\left(\frac{x\beta_{+}+\kappa q^{n}\beta_{-}}{\beta_{1,N}}\right)xG(x)k_{1,N}(x),\label{nd.23}
\end{equation}
\begin{equation}
k_{i,n+1}(x)=\epsilon q^{t_{i}-2n-1}\Gamma(n)\left(\frac{\beta_{1,i^{\prime}}}{\beta_{1,N}}\right)\left(\frac{x\beta_{+}+\epsilon q^{n}\beta_{-}}{\beta_{1,N}}\right)xG(x)k_{1,N}(x),\label{nd.24}
\end{equation}
\begin{equation}
k_{i,n+2}(x)=\epsilon q^{t_{i}-2n-1}\Gamma(n)\left(\frac{\beta_{1,i^{\prime}}}{\beta_{1,N}}\right)\left(\frac{x\beta_{+}-\epsilon q^{n}\beta_{-}}{\beta_{1,N}}\right)xG(x)k_{1,N}(x),\label{nd.25}
\end{equation}
where we have identified
\begin{eqnarray}
G(x) & = & \frac{q^{n-1}+\epsilon}{q^{n-1}+\epsilon x^{2}},\qquad\Gamma(n)=\frac{q^{n-1}+\epsilon}{q+1},\label{nd.26}
\end{eqnarray}
and 
\begin{equation}
\beta_{\pm}=\frac{1}{2}\left(\beta_{1,n+1}\pm\beta_{1,n+2}\right).
\end{equation}
Here we observe in Eqs.(\ref{nd.1}-\ref{nd.26}) the $\kappa$ dependency
inherited of the $R$-matrix and the new parameter $\epsilon=\pm1$,
besides the $\beta_{i,j}$ parameters.

\subsection{The block diagonal matrix elements\label{sub:Kd}}

Here we also identify a common structure for the diagonal matrix elements
$k_{i,i}(x)$ $(i\neq n+1,n+2)$ , namely
\begin{eqnarray}
k_{i,i}(x) & = & k_{i-1,i-1}(x)+\left(\frac{\beta_{i,i}-\beta_{i-1,i-1}}{\beta_{1,N}}\right)G(x)k_{1,N}(x),\nonumber \\
i & = & 2,...,n.\label{diag.1}
\end{eqnarray}
\begin{eqnarray}
k_{j,j}(x) & = & k_{j-1,j-1}(x)+\left(\frac{\beta_{j,j}-\beta_{j-1,j-1}}{\beta_{1,N}}\right)G(x)x^{2}k_{1,N}(x),\nonumber \\
j & = & n+4,...,N.\label{diag.2}
\end{eqnarray}
The central elements $k_{i,j}(x)$ $(i,j=n+1,n+2)$ satisfy a relation
slightly different from Eq.(\ref{diag.2}):
\begin{equation}
k_{n+2,n+2}(x)=k_{n+1,n+1}(x)+\left(\frac{\beta_{n+2,n+2}-\beta_{n+1,n+1}}{\beta_{1,N}}\right)xG(x)k_{1,N}(x),\label{cdiag.1}
\end{equation}
and
\begin{equation}
k_{n+2,n+1}(x)=k_{n+1,n+2}(x)+\left(\frac{\beta_{n+2,n+1}-\beta_{n+1,n+2}}{\beta_{1,N}}\right)xG(x)k_{1,N}(x).\label{cdiag.2}
\end{equation}

At this point we still have to find the remained matrix elements $k_{1,1}(x)$,
$k_{n+1,n+1}(x)$, $k_{n+1,n+2}(x)$ and $k_{n+3,n+3}(x)$ in terms
of $k_{1,N}(x)$. This issue proved to be a very difficult task, but
we managed to solve it completely as follows. 

From the non-diagonal elements presented above one can see that all
corresponding parameters given by Eq.(\ref{param.}) are determined
in terms of $\beta_{1,j}$ $(j=2,...,N)$ and $\beta_{2,1}$. Taking
into account the block diagonal terms, we still have to solve several
cumbersome algebraic equations with five unknown and $2N$ free parameters.
By inspection of these equations we can immediately see that the parameters
$\beta_{1,j\text{ }}$for $j=2,..,n-1$ and $\beta_{2,1}$ are determined
by 
\begin{equation}
\beta_{1,j}=\sigma(-1)^{j}\frac{\beta_{1,n}\beta_{1,n+3}}{\beta_{1,j^{\prime}}},\label{beta1j}
\end{equation}
\begin{equation}
\beta_{2,1}=-\sigma q^{3-2n}\Gamma(n)^{2}\frac{\beta_{1,n}\beta_{1,n+3}\beta_{1,N-1}}{\beta_{1,N}^{2}},\label{beta21}
\end{equation}
 where it is explicit a $n$-parity given by $\sigma=(-1)^{n}$. \ After
these computations we made the choice
\begin{equation}
k_{1,N}(x)=\frac{1}{2}\beta_{1,N}(x^{2}-1),\label{choice}
\end{equation}
in order to simplify our presentation. \ Here we note from the general
solution that $k_{1,N}(x)$ is an arbitrary function satisfying the
regularity condition $k_{1,N}(1)=0$. Therefore, the choice Eq.(\ref{choice})
does not imply any restriction as compared to the general case.

From this choice it follows appropriated expressions for $k_{1,1}(x)$
and $k_{n+3,n+3}(x)$:
\begin{eqnarray}
k_{1,1}(x) & = & \frac{\sqrt{q}\ \Gamma(n)G(x)}{2(q+1)q^{n}(x^{2}+1)\beta_{1,N}}\nonumber \\
 &  & \times\left\{ -\epsilon\sigma\sqrt{q}(x^{2}-\epsilon\sigma)\left[2(q^{n}+\sigma)+\epsilon(q-1)(x^{2}+\epsilon\sigma)\right]\beta_{1,n}\beta_{1,n+3}\right.\nonumber \\
 &  & -\left.\kappa(q+1)(q^{n}-\epsilon x^{2})\left[\left(q^{n}x^{2}-\kappa\right)\left(\beta_{+}^{2}+\beta_{-}^{2}\right)-\left(q^{n}x^{2}+\kappa\right)\left(\beta_{+}^{2}-\beta_{-}^{2}\right)\right]\right\} \nonumber \\
\label{k11}
\end{eqnarray}
and
\begin{eqnarray}
k_{n+3,n+3}(x) & = & \frac{\sqrt{q}\ \Gamma(n)G(x)x^{2}}{2(q+1)q^{n}(x^{2}+1)\beta_{1,N}}\nonumber \\
 &  & \times\left\{ \sqrt{q}(x^{2}-\epsilon\sigma)\left[2(q^{n}+\sigma)+\epsilon(q-1)(x^{2}+\epsilon\sigma)\right]\beta_{1,n}\beta_{1,n+3}\right.\nonumber \\
 &  & -\left.\kappa(q+1)(q^{n}-\epsilon x^{2})\left[\left(q^{n}-\kappa x^{2}\right)\left(\beta_{+}^{2}+\beta_{-}^{2}\right)-\left(q^{n}+\kappa x^{2}\right)\left(\beta_{+}^{2}-\beta_{-}^{2}\right)\right]\right\} .\nonumber \\
\label{kn3n3}
\end{eqnarray}
Now we have several reflection equations involving the diagonal parameters
$\beta_{i,i}$ \ $(i\neq n+1,n+2)$, which are solved by the recurrence
relations
\begin{eqnarray}
\beta_{i,i} & = & \beta_{i-1,i-1}+\sigma(-1)^{i}\Gamma(n)\left(\frac{q+1}{q^{n+1-i}}\right)\left(\frac{\beta_{1,n}\beta_{1,+3}}{\beta_{1,N}}\right),\nonumber \\
i & = & 2,...,n,\label{bii}
\end{eqnarray}
and
\begin{eqnarray}
\beta_{j,j} & = & \beta_{j-1,j-1}-\epsilon\sigma(-1)^{j}\Gamma(n)\left(\frac{q+1}{q^{N+1-j}}\right)\left(\frac{\beta_{1,n}\beta_{1,+3}}{\beta_{1,N}}\right),\nonumber \\
j & = & n+4,...,N,\label{bjj}
\end{eqnarray}
where $\beta_{1,1}$ and $\beta_{n+3,n+3}$ are determined by Eq.(\ref{param.}).

Finally, the central terms are determined by
\begin{eqnarray}
k_{n+1,n+1}(x) & = & \frac{\sqrt{q}\ \Gamma(n)G(x)}{4q^{n}(q+1)\beta_{1,N}}\left\{ (1-\epsilon\sigma)\sqrt{q}(q^{n}+\sigma x^{2})(x^{2}+1)\beta_{1,n}\beta_{1,n+3}\right.\nonumber \\
 &  & -\left.\kappa(q^{n}-\epsilon)(q+1)x^{2}\left[\left(q^{n}-\kappa\right)\left(\beta_{+}^{2}+\beta_{-}^{2}\right)-\left(q^{n}+\kappa\right)\left(\beta_{+}^{2}-\beta_{-}^{2}\right)\right]\right.\nonumber \\
 &  & -\left.2\left(\kappa\epsilon-1\right)q^{n}(q+1)x(x^{2}-1)\beta_{+}\beta_{-}\right\} ,\nonumber \\
\label{kn1n1}
\end{eqnarray}
\begin{eqnarray}
k_{n+1,n+2}(x) & = & \frac{\sqrt{q}\ \Gamma(n)G(x)}{4q^{n}(q+1)\beta_{1,N}}\left(\frac{x^{2}-1}{x^{2}+1}\right)\left\{ (1+\epsilon\sigma)\sqrt{q}(q^{n}-\sigma x^{2})(x^{2}-1)\beta_{1,n}\beta_{1,n+3}\right.\nonumber \\
 &  & +\left.\kappa(q^{n}+\epsilon)(q+1)x^{2}\left[\left(q^{n}+\kappa\right)\left(\beta_{+}^{2}+\beta_{-}^{2}\right)-\left(q^{n}-\kappa\right)\left(\beta_{+}^{2}-\beta_{-}^{2}\right)\right]\right.\nonumber \\
 &  & -\left.2\left(\kappa\epsilon+1\right)q^{n}(q+1)x(x^{2}+1)\beta_{+}\beta_{-}\right\} ,\nonumber \\
\label{kn1n2}
\end{eqnarray}
from which we can get the central parameters 
\begin{equation}
\beta_{n+2,n+1}=\beta_{n+1,n+2}+2\left(\kappa\epsilon+1\right)\sqrt{q}\ \Gamma(n)\left(\frac{\beta_{+}\beta_{-}}{\beta_{1,N}}\right),\label{bn2n1}
\end{equation}
\begin{equation}
\beta_{n+2,n+2}=\beta_{n+1,n+1}+2\left(\kappa\epsilon-1\right)\sqrt{q}\ \Gamma(n)\left(\frac{\beta_{+}\beta_{-}}{\beta_{1,N}}\right),\label{bn2n2}
\end{equation}
and then we can find the last parameter
\begin{eqnarray}
\beta_{1,N} & = & \frac{\sqrt{q}\ (q^{n}-\epsilon)\Gamma(n)}{4q^{n}(q+1)}\left\{ 2(1-\epsilon\sigma)\sqrt{q}\beta_{1,n}\beta_{1,n+3}\right.\nonumber \\
 &  & \left.-\kappa(q+1)\left[\left(q^{n}-\kappa\right)\left(\beta_{+}^{2}+\beta_{-}^{2}\right)-\left(q^{n}+\kappa\right)\left(\beta_{+}^{2}-\beta_{-}^{2}\right)\right]\right\} .\label{b1N}
\end{eqnarray}
At this point almost all equations are satisfied. For the remaining
equations be satisfied is necessary further to impose the constraint
\begin{equation}
(\nu-1)\beta_{+}\beta_{-}=0.\label{ceq}
\end{equation}

\section{Concluding remarks\label{sec:KC}}

From the results presented above we can identify 12 families of solutions
$S(\nu,\kappa,\epsilon)$ of the the multiparametric ${\cal {U}}_{q}[D_{n+1}^{(2)}]$
vertex model. For $\nu=1$ we obtain 4 families of solutions (regarding
the possibilities of $\kappa=\pm1$ and $\epsilon=\pm1$) characterized
by the $n+2$ free parameters $\beta_{1,n},...,\beta_{1,2n+1}$. For
$\nu=-1$ the solutions gets duplicated because of the constraint
Eq.(\ref{ceq}) and we obtain 8 more families of solutions (regarding
the possibilities of $\kappa=\pm1$, $\epsilon=\pm1$ and the two
possibilities, $\beta_{+}=0$ or $\beta_{-}=0$). In this case the
solution presents only $n+1$ free parameters, of course.

We highlight furthermore that the families $S(1,1,\epsilon)$ correspond
to the $K$-matrix solutions of the Jimbo's ${\cal U}_{q}[D_{n+1}^{(2)}]$
vertex model. The family $S(1,1,1)$ was already found by Lima-Santos
in \cite{LIM2}, while the family $S(1,1,-1)$ represents a new $K$-matrix
solution for this vertex model.

In the reference \cite{WM1} the baxterization of the representations
of the Birman-Wenzl-Murakami algebra was also used to produce multiparametric
solutions of the Yang-Baxter equation invariant by quantum superalgebras.
From these representations one can in principle construct new $K$-matrices
via our study. Thus, new open vertex models could be derived. It would
be interesting to know the type of open lattice models with both bosonic
and fermionic degrees of freedom that can be obtained. We hope to
report on this problem in a future publication.


\section*{Acknowledgements}

\addcontentsline{toc}{section}{Appendix A}The author Ricardo S.
Vieira thanks FAPESP (Fundação de Amparo à Pesquisa do Estado de São
Paulo) for financial support. The work of A. Lima-Santos has been
supported by CNPq-Brasil (Conselho Nacional de Desenvolvimento Científico
e Tecnológico) and FAPESP.

\section*{Appendix A: The block diagonal solutions\label{sec:Abd}}

\setcounter{equation}{0} \global\long\def\theequation{A.\arabic{equation}}

\addcontentsline{toc}{section}{Appendix B}

As was already mentioned in \cite{GM}, the ${\cal U}_{q}[D_{n+1}^{(2)}]$
vertex models have $n+1$ distinct $U(1)$ conserved charges, and
the $K$-matrix ansatz compatible with these symmetries is a block
diagonal structure.

Looking for the general solution of the corresponding reflection equation
we find that the only possible solution is obtained when the two recurrence
relations Eq.(\ref{diag.1}) and Eq.(\ref{diag.2}) are degenerated
into $k_{1,1}(x)$ and into $k_{N,N}(x)$, respectively. Thus, the
block diagonal structure has the form
\begin{equation}
K(x)={\rm diag}\left(k_{1,1}(x),\cdots,k_{1,1}(x),{\cal B}(x),k_{N,N}(x),\cdots,k_{N,N}(x)\right),\label{A.1}
\end{equation}
$\ $where ${\cal B}$ \ contains the central elements, 
\begin{equation}
{\cal B}(x)=\left(\begin{array}{c}
\begin{array}{cc}
k_{n+1,n+1}(x) & k_{n+1,n+2}(x)\\
k_{n+2,n+1}(x) & k_{n+2,n+2}(x)
\end{array}\end{array}\right).\label{A.2}
\end{equation}
The solution can be obtained by the same procedure described before
and in what follows we only quote the final results.

We have found two solutions for any value of $n$. \ The first solution
is given by
\begin{equation}
k_{N,N}(x)=\frac{(q^{n}-\kappa\nu)(x^{2}+1)+\beta_{n+1,n+2}(q^{n}+\kappa\nu)(x^{2}-1)}{(q^{n}-\kappa\nu)(x^{2}+1)-\beta_{n+1,n+2}(q^{n}+\kappa\nu)(x^{2}-1)}x^{2}k_{1,1}(x),\label{A.3}
\end{equation}
with central elements 
\begin{eqnarray}
k_{n+1,n+2}(x) & = & k_{n+2,n+1}(x)\nonumber \\
 & = & \frac{\beta_{n+1,n+2}x^{2}(x^{2}-1)(q^{2n}-1)k_{1,1}(x)}{(\kappa\nu x^{2}+q^{n})\left[(q^{n}-\kappa\nu)(x^{2}+1)-\beta_{n+1,n+2}(q^{n}+\kappa\nu)(x^{2}-1)\right]},\nonumber \\
\label{A.4}
\end{eqnarray}
\begin{eqnarray}
k_{n+1,n+1}(x) & = & \frac{x(x^{2}+1)\left[(q^{2n}-1)x+(x^{2}-1)\Sigma(n)\right]k_{1,1}(x)}{(\kappa\nu x^{2}+q^{n})\left[(q^{n}-\kappa\nu)(x^{2}+1)-\beta_{n+1,n+2}(q^{n}+\kappa\nu)(x^{2}-1)\right]},\nonumber \\
\label{A.5}
\end{eqnarray}
\begin{eqnarray}
k_{n+2,n+2}(x) & = & \frac{x(x^{2}+1)\left[(q^{2n}-1)x-(x^{2}-1)\Sigma(n)\right]k_{1,1}(x)}{(\kappa\nu x^{2}+q^{n})\left[(q^{n}-\kappa\nu)(x^{2}+1)-\beta_{n+1,n+2}(q^{n}+\kappa\nu)(x^{2}-1)\right]},\nonumber \\
\end{eqnarray}
where
\begin{equation}
\Sigma(n)=\epsilon\sqrt{\kappa\nu q^{n}\left[(q^{n}+\kappa\nu)^{2}\beta_{n+1,n+2}^{2}-(q^{n}-\kappa\nu)^{2}\right]}.\label{A.6}
\end{equation}
The parameter $\epsilon=\pm1$ indicates the existence of two conjugated
solutions and $k_{1,1}(x)$ can be any function that satisfies the
regularity condition. 

Moreover we notice that these solutions degenerate into four diagonal
solutions when $\beta_{n+1,n+2}=0$, namely
\begin{equation}
k_{n,n}(x)=\cdots=k_{2,2}(x)=k_{1,1}(x),\label{A.7}
\end{equation}
\begin{equation}
k_{n+1,n+1}(x)=\frac{(q^{2n}-1)x+\epsilon\sqrt{-\kappa\nu q^{n}\left(q^{n}-\kappa\nu\right)^{2}}\left(x^{2}-1\right)}{\left(q^{n}+\kappa\nu x^{2}\right)\left(q^{n}-\kappa\nu\right)}xk_{1,1}(x),\label{A.8}
\end{equation}
\begin{equation}
k_{n+2,n+2}(x)=\frac{(q^{2n}-1)x-\epsilon\sqrt{-\kappa\nu q^{n}\left(q^{n}-\kappa\nu\right)^{2}}\left(x^{2}-1\right)}{\left(q^{n}+\kappa\nu x^{2}\right)\left(q^{n}-\kappa\nu\right)}xk_{1,1}(x),\label{A.9}
\end{equation}

\begin{equation}
k_{N,N}(x)=\cdots=k_{n+3,n+3}(x)=x^{2}k_{1,1}(x).\label{A.10}
\end{equation}
Here we note that the trivial diagonal solution ($K^{-}(x)={\bf I}$
and $K^{+}(x)=V^{t}V$) does not holds for this system \cite{MEZ}.

For the second family of block diagonal solutions, in order to simplify
our presentation, we made the choice
\begin{equation}
k_{n+1,n+1}(x)=k_{n+2,n+2}(x)=\frac{1}{2}x(x^{2}+1).\label{A.11}
\end{equation}
With this choice, the remained matrix elements different from zero
are given by
\begin{eqnarray}
k_{1,1}(x) & = & \frac{1}{2}\frac{(q^{n}-\kappa\nu x^{2})}{x(q^{n}-\kappa\nu)^{2}}\left\{ -(x^{2}-1)\left[(q^{n}+\kappa\nu)\beta_{n+1,n+2}+2\Omega(n)\right]\right.\nonumber \\
 &  & +\left.(x^{2}+1)(q^{n}-\kappa\nu)\right\} ,\label{A.12}
\end{eqnarray}

\begin{eqnarray}
k_{N,N}(x) & = & \frac{1}{2}\frac{x(q^{n}-\kappa\nu x^{2})}{(q^{n}-\kappa\nu)^{2}}\left\{ (x^{2}-1)\left[(q^{n}+\kappa\nu)\beta_{n+1,n+2}+2\Omega(n)\right]\right.\nonumber \\
 &  & +\left.(x^{2}+1)(q^{n}-\kappa\nu)\right\} ,\label{A.13}
\end{eqnarray}
and
\begin{eqnarray}
k_{n+1,n+2}(x) & = & \frac{1}{2}\frac{(x^{2}-1)}{(q^{n}-\kappa\nu)^{2}}\left\{ \beta_{n+1,n+2}\left[x(q^{n}+\kappa\nu)^{2}-2\kappa\nu q^{n}(x^{2}+1)\right]\right.\nonumber \\
 &  & -\left.(q^{n}+\kappa\nu)\Omega(n)(x-1)^{2}\right\} ,\label{A.14}
\end{eqnarray}
\begin{eqnarray}
k_{n+2,n+1}(x) & = & \frac{1}{2}\frac{(x^{2}-1)}{(q^{n}-\kappa\nu)^{2}}\left\{ \beta_{n+1,n+2}\left[x(q^{n}+\kappa\nu)^{2}+2\kappa\nu q^{n}(x^{2}+1)\right]\right.\nonumber \\
 &  & +\left.(q^{n}+\kappa\nu)\Omega(n)(x+1)^{2}\right\} .\label{A.15}
\end{eqnarray}
where
\begin{equation}
\Omega(n)=\epsilon\sqrt{\kappa\nu q^{n}\left(\beta_{n+1,n+2}^{2}-1\right)}.\label{A.16}
\end{equation}
In contrast to the first solution there is no way to derive a diagonal
solution from these four conjugated solutions.

\section*{Appendix B: The multiparametric ${\cal U}_{q}[D_{2}^{(2)}]$ vertex
model\label{sec:AD2}}

\setcounter{equation}{0} \global\long\def\theequation{B.\arabic{equation}}

\addcontentsline{toc}{section}{Appendix B}

As mentioned in the section \ref{sec:K}, the most general $K$-matrix
associated with the multiparametric ${\cal U}_{q}[D_{n+1}^{(2)}]$
vertex model has $n+2$ free parameters, namely $\beta_{1,n}$, $\beta_{1,n+1}$,
$...,\beta_{1,N-1}$. It is interesting to notice however that only
$\beta_{1,n}$, $\beta_{1,n+1}$, $\beta_{1,n+2}$, $\beta_{1,n+3}$
and $\beta_{1,N-1}$ appears explicitly on the solution. For $n=1$,
however, an ambiguity arises due to this choice of writing the solution.
Indeed, one can realize that in this case some of those free parameters
get confused with others (for instance, we would have $\beta_{1,n+1}=\beta_{1,N-1}$
and $\beta_{1,n+2}=\beta_{1,N}$). Therefore, it is clear that the
${\cal U}_{q}[D_{2}^{(2)}]$ vertex model should be treated separately,
in order to one find its general solution. 

Proceeding in this way, we shall see that the elements out of the
diagonals can actually be read direct from the equations (\ref{nd.1}-\ref{nd.25}),
of the solution presented in the section \ref{sec:K}, if we replace
there $\beta_{1,N-1}$ by $\left(\beta_{+}-\kappa\beta_{-}\right)/\sqrt{q}$
and make $\epsilon=1$ (the choice $\epsilon=-1$ leads to the trivial
solution $K(x)=0$). This means that the elements of the first row
are given by 
\begin{eqnarray}
k_{1,2}(x) & = & \left(\beta_{+}+x\beta_{-}\right)G(x)\frac{k_{1,4}(x)}{\beta_{1,4}},\\
k_{1,3}(x) & = & \left(\beta_{+}-x\beta_{-}\right)G(x)\frac{k_{1,4}(x)}{\beta_{1,4}},
\end{eqnarray}
and the elements of the first column, are given by 
\begin{eqnarray}
k_{2,1}(x) & = & \left(\frac{\beta_{+}-\kappa x\beta_{-}}{\beta_{+}-\kappa\beta_{-}}\right)\frac{\beta_{2,1}}{\beta_{1,4}}G(x)k_{1,4}(x),\\
k_{3,1}(x) & = & \left(\frac{\beta_{+}+\kappa x\beta_{-}}{\beta_{+}-\kappa\beta_{-}}\right)\frac{\beta_{2,1}}{\beta_{1,4}}G(x)k_{1,4}(x).
\end{eqnarray}
 For the elements of the last row we have 
\begin{eqnarray}
k_{2,4}(x) & = & \left(x\beta_{+}-\kappa q\beta_{-}\right)\frac{xG(x)}{\sqrt{q}}\frac{k_{1,4}(x)}{\beta_{1,4}},\\
k_{3,4}(x) & = & \left(x\beta_{+}-\kappa q\beta_{-}\right)\frac{xG(x)}{\sqrt{q}}\frac{k_{1,4}(x)}{\beta_{1,4}},
\end{eqnarray}
 and, on the last column, 
\begin{eqnarray}
k_{4,1}(x) & = & \frac{\beta_{4,1}}{\beta_{1,4}}k_{1,4}(x),\\
k_{4,2}(x) & = & \left(\frac{x\beta_{+}+q\beta_{-}}{\beta_{+}-\kappa\beta_{-}}\right)\frac{\beta_{2,1}}{\beta_{1,4}}\frac{xG(x)}{\sqrt{q}}k_{1,4}(x),\\
k_{4,3}(x) & = & \left(\frac{x\beta_{+}-q\beta_{-}}{\beta_{+}-\kappa\beta_{-}}\right)\frac{\beta_{2,1}}{\beta_{1,4}}\frac{xG(x)}{\sqrt{q}}k_{1,4}(x),
\end{eqnarray}
where, we should remember, for $n=1$ and $\epsilon=1$, we get 
\begin{equation}
G(x)=\frac{2}{x^{2}+1},\qquad\Gamma(1)=\frac{2}{q+1}.
\end{equation}

Furthermore, on the main diagonal we have

\begin{eqnarray}
k_{1,1}(x) & = & 1+\frac{\Gamma(1)G(x)k_{1,4}(x)}{4\sqrt{q}\ \left(x^{2}+1\right)\beta_{1,4}^{2}}\left\{ \left(x^{2}-1\right)\left(q+1\right)^{2}\left(\beta_{+}^{2}+\kappa\beta_{-}^{2}\right)\right.\nonumber \\
 &  & -\left.\left[\left(x^{2}-1\right)\left(q^{2}-1\right)+4q\left(x^{2}+1\right)\right]\left(\beta_{+}^{2}-\kappa\beta_{-}^{2}\right)\right\} ,\nonumber \\
\end{eqnarray}

and
\begin{eqnarray}
k_{4,4}(x) & = & x^{2}+\frac{\Gamma(1)G(x)x^{2}k_{1,4}(x)}{4\sqrt{q}\ (x^{2}+1)\beta_{1,4}^{2}}\nonumber \\
 &  & \times\left\{ \left(q+1\right)\left[\left(x^{2}q-1\right)-3\left(x^{2}-q\right)\right]\left(\beta_{-}^{2}\kappa+\beta_{+}^{2}\right)\right.\nonumber \\
 &  & -\left.\left[\left(q^{2}x^{2}+1\right)+3\left(x^{2}+q^{2}\right)\right]\left(\beta_{+}^{2}-\beta_{-}^{2}\kappa\right)\right\} .\nonumber \\
\end{eqnarray}

For the central block we can verify that the central matrix elements
are still given by the Eqs.(\ref{cdiag.1},\ref{cdiag.2}),
\begin{eqnarray}
k_{3,2}(x) & = & k_{2,3}(x)+\left(\frac{\beta_{3,2}-\beta_{2,3}}{\beta_{1,4}}\right)G(x)xk_{1,4}(x),\\
k_{3,3}(x) & = & k_{2,2}(x)+\left(\frac{\beta_{3,3}-\beta_{2,2}}{\beta_{1,4}}\right)G(x)xk_{1,4}(x),
\end{eqnarray}
but now we have 
\begin{eqnarray}
k_{2,2}(x) & = & \frac{q-x^{2}}{q-1}+\frac{\Gamma(1)G(x)k_{1,4}(x)}{4\sqrt{q}\ \beta_{1,4}^{2}}\nonumber \\
 &  & \times\left\{ \kappa\left(x^{2}+q\right)\left[(q+\kappa)(\beta_{+}^{2}-\beta_{-}^{2})-(q-\kappa)(\beta_{+}^{2}+\beta_{-}^{2})\right]\right.\nonumber \\
 &  & -\left.4qx\left(\kappa-1\right)\beta_{+}\beta_{-}\right\} ,\label{d22.8}
\end{eqnarray}
and
\begin{eqnarray}
k_{2,3}(x) & = & \frac{\Gamma(1)G(x)k_{1,4}(x)}{4\sqrt{q}\ (x^{2}+1)\beta_{1,4}^{2}}\nonumber \\
 &  & \times\left\{ \kappa x(q+1)\left[(q+\kappa)(\beta_{+}^{2}+\beta_{-}^{2})-2(q-\kappa)(\beta_{+}^{2}-\beta_{-}^{2})\right]\right.\nonumber \\
 &  & -\left.4q(x^{2}+1)(\kappa+1)\beta_{+}\beta_{-}\right\} .\label{d22.9}
\end{eqnarray}
After we get the parameters
\begin{equation}
\beta_{{3,2}}=\beta_{{2,3}}+2\left(\kappa+1\right)\sqrt{q}\ \Gamma(1)\left(\frac{\beta_{-}\beta_{+}}{\beta_{{1,4}}}\right),\label{d22.13}
\end{equation}
\begin{equation}
\beta_{{3,3}}=\beta_{{2,2}}+2\left(\kappa-1\right)\sqrt{q}\ \Gamma(1)\left(\frac{\beta_{-}\beta_{+}}{\beta_{{1,4}}}\right),\label{d22.14}
\end{equation}
and 
\begin{equation}
\beta_{2,1}=\frac{2\sqrt{q}}{q-1}\left(\frac{\beta_{+}+\beta_{-}\kappa}{\beta_{1,4}}\right)\left[1-\frac{\kappa\sqrt{q}\ \Gamma(1)\left(q-1\right)}{2}\left(\frac{\beta_{+}^{2}-\beta_{-}^{2}}{\beta_{1,4}}\right)\right],\label{d22.12}
\end{equation}
\begin{equation}
\beta_{4,1}=\frac{\beta_{1,4}\beta_{2,1}^{2}}{\left(\beta_{+}-\kappa\beta_{-}\right)^{2}}.\label{d22.15}
\end{equation}

The remained reflection equations are also constrained by the Eq.(\ref{ceq})
and then we get a solution with 3 free parameters ($\beta_{1,2},\beta_{1,3}$
and $\beta_{1,4}$) for $\nu=1$ and with only 2 free parameters for
$\nu=-1$. 

Finally, we mention that the block diagonal and the diagonal solutions
for the multiparametric ${\cal U}_{q}[D_{2}^{(2)}]$ vertex model
can be obtained straightforward from the appendix $A$, taking into
account $n=1$.

\section*{Appendix C: Special solutions\label{sec:As}}

\setcounter{equation}{0} \global\long\def\theequation{C.\arabic{equation}}

\addcontentsline{toc}{section}{Appendix C}

Again, from the constraint equations Eq.(\ref{ceq}) we look at the
possibility $\beta_{+}=\beta_{-}=0$. \ In this case our multiparametric
$K$-matrix has the form
\begin{equation}
K^{-}(x)=\left(\begin{array}{ccc}
{\bf A}(x) & {\bf 0} & {\bf B}(x)\\
{\bf 0} & {\bf E}(x) & {\bf 0}\\
{\bf C}(x) & {\bf 0} & {\bf D}(x)
\end{array}\right),\label{KC}
\end{equation}
where the ${\bf 0}^{\prime}s$ are null matrices, ${\bf A}(x)$, ${\bf B}(x)$,${\bf C}(x)$,
${\bf D}(x)$ are $n$ by $n$ matrices and 
\begin{equation}
{\bf E}(x)=\left(\begin{array}{cc}
k_{n+1,n+1}(x) & 0\\
0 & k_{n+2,n+2}(x)
\end{array}\right).
\end{equation}
This results in a $n$ free parameter solution. However, from this
special form of the $K$-matrix we have found new solutions, in analogy
with appearance of new $R$-matrices satisfied the Yang-Baxter equations.

Our first special solution does not depend on the parameters $\kappa$
and $\nu$ and its non zero matrix elements are only in the diagonals.
Making $k_{1,N}(x)=\beta_{1,N}(x^{2}-1)/2$, the main diagonal block
matrices are
\begin{equation}
{\bf A}(x)={\bf I}_{n\times n},\quad{\bf E}(x)=\left(\begin{array}{cc}
\frac{q-x^{2}}{q-1} & 0\\
0 & \frac{q-x^{2}}{q-1}
\end{array}\right),\text{\quad}{\bf D}(x)=x^{2}{\bf I}_{n\times n}
\end{equation}
where ${\bf I}_{n\times n}$ is the $n$ by $n$ unity matrix. \ The
minor diagonal matrix elements are 
\begin{eqnarray}
({\bf B})_{i,j} & = & \frac{1}{2}\beta_{i,j}(x^{2}-1)\delta_{j,i^{\prime},}\quad i=1,2,...,n,\nonumber \\
({\bf C})_{i,j} & = & \frac{1}{2}\beta_{i,j}(x^{2}-1)\delta_{j,i^{\prime}},\quad i=n+3,...,N,
\end{eqnarray}
where the parameters $\beta_{i,j}$ are related by 
\begin{equation}
\beta_{i,i^{\prime}}\beta_{i^{\prime},i}=\frac{4q}{(q-1)^{2}},\qquad i=1,2,...,n.
\end{equation}
It means that we have found a new $n$ free parameter solution which
has the form of the letter X.

A second special solution was obtained when $n=2$, that is, for the
multiparametric ${\cal U}_{q}[D_{3}^{(2)}]$ vertex model. In this
case follows from the Eq.(\ref{KC}) that the $K$-matrix has the
form 
\begin{equation}
K^{-}(x)=\begin{pmatrix}\begin{array}{cc}
k_{1,1}(x) & k_{1,2}(x)\\
k_{2,1}(x) & k_{2,2}(x)
\end{array} & 0 & 0 & \begin{array}{cc}
k_{1,5}(x) & k_{1,6}(x)\\
k_{2,5}(x) & k_{2,6}(x)
\end{array}\\
0 & k_{3,3}(x) & 0 & 0\\
0 & 0 & k_{4,4}(x) & 0\\
\begin{array}{cc}
k_{5,1}(x) & k_{5,2}(x)\\
k_{6,1}(x) & k_{6,2}(x)
\end{array} & 0 & 0 & \begin{array}{cc}
k_{5,5}(x) & k_{5,6}(x)\\
k_{6,5}(x) & k_{6,6}(x)
\end{array}
\end{pmatrix}.
\end{equation}
By solving this problem separately we can see that the elements out
of the diagonals can be read directly from the solution presented
on the section (\ref{sec:K}), provided we set there $n=2$ and $\epsilon=-1$.
This means we have 
\begin{eqnarray}
k_{1,2}(x) & = & \frac{\beta_{1,2}G(x)k_{1,6}(x)}{\beta_{1,6}},\\
k_{2,1}(x) & = & \frac{\beta_{2,1}G(x)k_{1,6}(x)}{\beta_{1,6}},\\
k_{1,5}(x) & = & \frac{\beta_{1,5}G(x)k_{1,6}(x)}{\beta_{1,6}},\\
k_{2,6}(x) & = & -\frac{\beta_{1,5}x^{2}G(x)k_{1,6}(x)}{q\beta_{1,6}},
\end{eqnarray}
\begin{eqnarray}
k_{5,1}(x) & = & \frac{q\beta_{1,2}\beta_{2,1}G(x)k_{1,6}(x)}{\beta_{1,5}\beta_{1,6}},\\
k_{6,2}(x) & = & -\frac{\beta_{1,2}\beta_{2,1}x^{2}G(x)k_{1,6}(x)}{\beta_{1,5}\beta_{1,6}},\\
k_{5,6}(x) & = & -\frac{\beta_{1,2}x^{2}G(x)k_{1,6}(x)}{\beta_{1,6}},\\
k_{6,5}(x) & = & -\frac{\beta_{2,1}x^{2}G(x)k_{1,6}(x)}{\beta_{1,6}},
\end{eqnarray}
 where now $\beta_{2,1}$ is given by 
\begin{equation}
\beta_{2,1}=\left(\frac{\beta_{1,2}\beta_{1,5}}{q\beta_{1,6}}-\frac{2}{q-1}\right)\frac{\beta_{1,5}}{\beta_{1,6}}.
\end{equation}

On the minor diagonal, we have, 
\begin{eqnarray}
k_{2,5}(x) & = & -\frac{\beta_{2,1}k_{1,6}(x)}{\beta_{1,2}},\\
k_{5,2}(x) & = & -\frac{q\beta_{1,2}\beta_{2,1}k_{1,6}(x)}{\beta_{1,2}\beta_{1,5}^{2}},\\
k_{6,1}(x) & = & -\frac{q\beta_{2,1}^{2}k_{1,6}(x)}{\beta_{1,5}^{2}}.
\end{eqnarray}
For the elements of the principal diagonal, the ansatz Eqs.(\ref{diag.1},\ref{diag.2})
still holds, so that we have 
\begin{eqnarray}
k_{2,2}(x) & = & k_{1,1}(x)+\left(\frac{\beta_{2,2}-\beta_{1,1}}{\beta_{1,N}}\right)G(x)k_{1,N}(x),\\
k_{6,6}(x) & = & k_{5,5}(x)+\left(\frac{\beta_{6,6}-\beta_{5,5}}{\beta_{1,N}}\right)x^{2}G(x)k_{1,N}(x),
\end{eqnarray}
 where now 
\begin{eqnarray}
k_{2,2}(x) & = & 1+\frac{\beta_{1,2}\beta_{1,5}G(x)k_{1,6}(x)}{\beta_{1,6}^{2}},\\
k_{5,5}(x) & = & x^{2}+\frac{\beta_{1,2}\beta_{1,5}x^{2}G(x)k_{1,6}(x)}{\beta_{1,6}^{2}}.
\end{eqnarray}
Finally, for the central block we have 
\begin{equation}
k_{4,4}(x)=k_{3,3}(x)=\frac{1}{q-1}+\left(\frac{q^{2}-x^{2}}{q-x^{2}}\right)\frac{\beta_{1,2}\beta_{1,5}k_{1,6}(x)}{q\beta_{1,6}^{2}}.
\end{equation}
In this way we obtain a solution which is valid to any value of $\kappa$
and $\nu$ and which is characterized by the free parameters $\beta_{1,2}$,
$\beta_{1,5}$ and $\beta_{1,6}$. 

To conclude we mention that this special solution was suggested by
the existence of the ``almost unitary'' solution 
\begin{equation}
K^{-}(x)={\rm diag}(x^{-2},1,1,1,x^{2}),\label{A.17}
\end{equation}
previously presented in \cite{GM}, which is valid only to this particular
model as well.

\setcounter{equation}{0} \global\long\def\theequation{B.\arabic{equation}}

\end{document}